\documentclass[12pt]{article}
\def\ii{\'{\i}}
\def\beq{\begin{equation}}
\def\eeq{\end{equation}}
\def\beqa{\begin{eqnarray}}
\def\eeqa{\end{eqnarray}}
\def\ban{\begin{eqnarray*}}
\def\ean{\end{eqnarray*}}
\def\bi{\begin{itemize}}
\def\ei{\end{itemize}}

\def\d{\mbox{d}}
\def\den{(\hat{\rho})}

\begin{document}

\begin{center}
{\bf On the density dependent hadron field theory at finite temperature \\
and its thermodynamical consistency}
\end{center}

\begin{center}
{\it S.S. Avancini $^1$, M. E. Bracco $^2$, M. Chiapparini $^2$
and D.P.Menezes $^1$}\\
{\it $^1$  Depto de F\'{\i}sica - CFM -
  Universidade Federal de Santa Catarina}\\
{\it Florian\'opolis - SC - CP. 476 - CEP 88.040 - 900 - Brazil}\\
{\it $^2$ Instituto de F\ii sica - Universidade do Estado do Rio de Janeiro}\\
{\it Rua S\~ao Francisco Xavier 524 - Maracan\~a - CEP 20559-900}\\
{\it Rio de Janeiro - RJ - Brazil}\\
\end{center}

\vspace{0.50cm}

\begin{abstract}
In this work we study in a formal way
the density dependent hadron field theory at finite temperature for
nuclear matter. The thermodynamical potential and related quantities, as
energy density and pressure are derived in two different ways. We first
obtain the thermodynamical potential from the grand partition function, where
the Hamiltonian depends on the density operator and is truncated at first
order. We then reobtain the thermodynamical potential by calculating
explicitly the energy density in a Thomas-Fermi approximation and considering
the entropy of a fermi gas. The distribution functions for particles and
antiparticles are the output of the minimization of the thermodynamical
potential.
It is shown that in the mean field theory the thermodynamical
consistency is achieved. The connection with
effective chiral lagrangians with Brown-Rho scaling is discussed.
\end{abstract}

\vspace{0.50cm}

PACS number(s): {\bf 21.65.+f, 21.30.-x, 25.70.-z}

\section{Introduction }
Recently effective field theories have been often used in order to
describe hadron properties since the solutions for the fundamental
theory of strong interactions, the Quantum Chromo Dynamics (QCD),
are still an unreachable problem at the hadron physics energy
scale. The study of  nuclear matter and finite nuclei properties
at finite temperature has become an important problem since a
large variety of data, where matter is being tested at extreme
conditions of density, pressure and non-zero temperature, are
becoming available in the modern experimental facilities which are
already operational.

The motivation for the formulation of Density Dependent Hadron Field Theory
(DDHFT) \cite{flw,lf} may be linked with the relativistic
Dirac-Brueckner Hartree-Fock (DBHF) theory \cite{DB}.
Recently, one approach that
has being used \cite{tw,ring1} is to consider the DBHF
 calculations for nuclear matter only as a guide for a
 suitable parametrization of density dependence of the meson-baryon
 coupling operators. In fact, the parameters are adjusted in
 order to describe the nuclear matter and some finite nuclei properties, as
 it is done in the usual Walecka model parametrizations \cite{sw}.
 In the present work we discuss how to treat the nuclear matter in the
 DDHFT at finite temperature and show that, in the mean field approximation
 (MFA), the thermodynamical consistency is achieved. This is in line
 with previous calculations done on the same subject, where
 thermodynamical consistency was assumed \cite{ds}. We also show
 that the mean field approximation is equivalent to the
 Thomas-Fermi approximation (TFA).
We follow two different methods in obtaining
our results. We first start from the calculation of the grand partition
function in terms of the density dependent Hamiltonian, which is truncated at
first order. Once the thermodynamical potential is found, all other
thermodynamical quantities can be obtained from it. In the second approach, we
start by writing the thermodynamical potential in terms of the energy density
calculated in the Thomas-Fermi approximation and the entropy of a fermi gas.
The distribution functions for particles and antiparticles are the result of
the minimization of the thermodynamical potential.

Furthermore, there is a
 class of  models, including those based on effective chiral
 lagrangians endowed with the Brown-Rho scaling, to which our formalism
 and our conclusions can be easily extended. This point is discussed later
in this paper.

The paper is organized as follows: in section 2 the DDHFT is presented.
In section 3 a mean field approximation is performed and some termodynamic
quantities are calculated. In section 4 the Thomas-Fermi approximation is
used and the results are compared with the ones obtained in section 3. Finally,
in the last section the conclusions are drawn.

\section{Density Dependent Hadron Field Theory Formalism }
We start from the density dependent hadron field theory Lagrangian
density of the well known Walecka-type models \cite{flw,lf,sw}:
\beqa
{\cal L}&=&\bar \psi\left[
\gamma_\mu\left(i\partial^{\mu}-\Gamma_\omega \den \omega^{\mu}-
\frac{\Gamma_{\rho}\den  }{2} \vec{\tau} \cdot \vec{\rho}^{~\mu} -
\frac{e}{2}\left(1+\tau_3 \right)  A^{\mu} \right)
-\left(M-\Gamma_\phi\den \phi\right)\right]\psi \nonumber \\
&&+\frac{1}{2}\left(\partial_{\mu}\phi\partial^{\mu}\phi -m_\phi^2
\phi^2\right) -\frac{1}{4}\omega_{\mu\nu}\omega^{\mu\nu}+\frac{1}{2}
m_\omega^2
\omega_{\mu}\omega^{\mu}  \nonumber \\
&&-\frac{1}{4}\vec\rho_{\mu\nu}\cdot\vec
\rho^{\mu\nu}+\frac{1}{2} m_\rho^2 \vec\rho_{\mu}\cdot \vec
\rho^{\mu} -\frac{1}{4}F_{\mu\nu} F^{\mu\nu}\;, \label{lag1}
\eeqa
where $\phi$, $\omega^\mu$, $\vec{\rho^\mu}$ and $A^{\mu}$ are the
scalar-isoscalar, vector-isoscalar and vector-isovector meson
fields and the photon field respectively,
$\omega_{\mu\nu}=\partial_{\mu}\omega_{\nu}-\partial_{\nu}\omega_{\mu}$,
$F_{\mu\nu}=\partial_{\mu}A_{\nu}-\partial_{\nu}A_{\mu}$, and
$\vec \rho_{\mu\nu}=\partial_{\mu}\vec \rho_{\nu}-\partial_{\nu}
\vec \rho_{\mu} - \Gamma_\rho \left(\vec \rho_\mu \times \vec
\rho_\nu\right)$, $M$ is the nucleon mass, $m_\phi$, $m_\omega$,
$m_\rho$ are the masses of the mesons and $\tau_3 = -1$ ($\tau_3 =
1$) for neutrons (protons). The nucleon-meson density dependent
coupling operators
 $\Gamma_\phi\den$, $\Gamma_\omega\den$, $\Gamma_\rho\den$
 are taken to be dependent on a Lorentz scalar functional,
 $\hat\rho\left(\bar\psi ,\psi\right)$. In what follows we assume the vector
 density dependence description \cite{flw}, i. e.,
 ${\hat\rho}^2=j^\mu j_\mu$, where $j_{\mu}=\bar \psi \gamma_\mu \psi$.
 Actually this has been the functional dependence most often
used in recent applications \cite{tw,ring1,ditoro}. Notice that
other possibilities for the functional dependencies of the
coupling operators may be considered as well, for example,
$\Gamma_\phi$ has been taken to be a function of $\hat{\rho}_s
=\bar \psi \psi$ in \cite{flw}.

In this work our emphasis is to consider hadronic models in the context of
the Walecka model. However, our formalism can be applied to a larger class of
systems, i.e., any Lagrangian density of the type
${\cal L}={\cal L}_F + {\cal L}_B + {\cal L}_{int}$, with
${\cal L}_F = \bar \psi\left(i \gamma^{\mu} \partial_{\mu} -M\right) \psi$
corresponding
to the fermion sector, ${\cal L}_B$ to the free bosons and ${\cal L}_{int}$
being any interaction which can written as
${\cal L}_{int} = \Gamma_{\alpha}(\hat \rho)\bar \psi
\Theta_{\alpha}\psi V^{\alpha}$, the summation running over the bosons,
$\Theta_{\alpha}$
containing Dirac and/or isospin matrices and $V^{\alpha}$ being any boson field
(scalar, vector-isovector, vector-isoscalar, etc).

As usual, the field equations of motion follow from the
Euler-Lagrange equations. Here some care has to be taken since the
coupling operators are now dependent on the baryon fields $\bar
\psi$ and $\psi$ through $\hat \rho$. Hence, if $\xi$ stands for a
generic field, the Euler-Lagrange equations are:
\beqa
\frac{\partial \cal{L}}{\partial \xi}
-\partial_\mu\left(\frac{\partial \cal{L}}{\partial \partial_\mu
\xi}\right) + \frac{\partial
\cal{L}}{\partial \hat\rho}\frac{\partial \hat\rho}{\partial \xi}
=0\;. \label{EEL}
\eeqa
When the partial derivatives
of $\cal{L}$ are performed relatively to the fields $\bar \psi$
and $\psi$, they yield extra terms due to the functional
dependence of the coupling
 operators in $\hat\rho(\bar\psi,\psi)$. These new terms are absent in the
usual Quantum Hadrod
Dynamic (QHD) model \cite{sw}. So, the equations of
motion for the fields read:
\beqa
\left(\partial_\mu\partial^{\mu} + m_{\phi}^2\right)\phi &=&
 \Gamma_\phi\den \bar \psi \psi\;, \label{PHI} \\
\partial_{\nu} \omega^{\mu\nu} + m_{\omega}^2 \omega^{\mu} &=&
 \Gamma_{\omega}\den \bar \psi \gamma^{\mu} \psi\;, \label{OME}\\
\partial_{\nu} {\vec{\rho}}^{~\mu\nu} + m_\rho^2
\vec{\rho}^{~\mu} &=&
\frac{\Gamma_\rho\den}{2} \bar \psi \vec{\tau} \gamma^{\mu}
\psi\;, \label{RHO}\\
\partial_{\nu} F^{\mu\nu} &=& \frac{e}{2}\bar \psi
\left(1+\tau_3 \right)\gamma^{\mu} \psi, \label{EM}\\
\left[ \gamma_{\mu}\left(i\partial^\mu -\Sigma^{\mu}\den\right) -M^{\ast}
\right] \psi&=&0\;, \label{DIRAC}
\eeqa
where $M^{\ast}=M-\Gamma_\phi\den\phi$. Notice that in the
equation of motion for the baryon field $\psi$ the vector
self-energy consists of two terms, $\Sigma_{\mu}$ =
$\Sigma^{(0)}_\mu$ + $\Sigma^{R}_{\mu}$, where:
\beqa
 \Sigma^{(0)}_\mu & = &\Gamma_{\omega}\den \omega_\mu
+\frac{\Gamma_{\rho}\den}{2} \vec{\tau}\cdot \vec{\rho}_{\mu}
+ \frac{e}{2} \left(1+\tau_3 \right) A_{\mu}\;, \\
\Sigma^{R}_{\mu}&=&\left( \frac{\partial \Gamma_{\omega}\den}{\partial
\hat{\rho}} \omega^\nu j_\nu  + \frac{1}{2} \frac{\partial
\Gamma_\rho\den}{\partial \hat{\rho}} \bar \psi \gamma^\nu
\vec{\tau}\cdot \vec{\rho}_\nu \psi   - \frac{\partial
\Gamma_\phi\den}{\partial \hat{\rho}} \bar \psi \psi \phi\right)
u_\mu \;, \label{REAR}
\eeqa
where $\Sigma^{(0)}_\mu$ is the usual
vector self-energy, $\hat\rho u_\mu=j_\mu$ with $u^2=1$ and, as a
result of the derivative of the Lagrangian with respect to
$\hat{\rho}$ a new term appears, $\Sigma^{R}_{\mu}$, which  is
called {\em rearrangement self-energy} and will play an essential r\^ole
in the applications of the theory. This term guarantees the
thermodynamical consistency and the energy-momentum conservation,
i.e., $\partial_\mu {\cal T}^{\mu\nu}=0$ of the density dependent
effective models, where the energy-momentum tensor is given by \cite{flw}:
\beqa
{\cal T}^{\mu \nu} &=& -g^{\mu \nu} \left[ \bar \psi
\gamma^{\lambda} \Sigma^{R}_{\lambda} \psi
 +\frac{1}{2} ( \partial_{\lambda}\phi\partial^{\lambda}
 \phi -m_\phi^2 \phi^2 )
-\frac{1}{4}\omega_{\lambda\eta}\omega^{\lambda\eta}+\frac{1}{2}
m_\omega^2
\omega_{\lambda}\omega^{\lambda} \right.   \nonumber  \\
&& \left. -\frac{1}{4}\vec{\rho}_{\lambda\eta}\cdot
 \vec{\rho}^{\lambda\eta}+\frac{1}{2} m_\rho^2 \vec{\rho}_{\lambda}\cdot
\vec \rho^{\lambda} -\frac{1}{4}F_{\lambda\eta} F^{\lambda\eta}
\right] \nonumber \\
&&+\bar \psi i \gamma^{\mu} \partial^{\nu} \psi +
\partial^{\mu}\phi\partial^{\nu}\phi + \partial^{\nu}
\omega_{\lambda} \omega^{\lambda\mu} + \partial^{\nu}
\vec{\rho}_{\lambda}\cdot \vec{\rho}^{\lambda \mu} +
\partial^{\nu} A_{\lambda} F^{\lambda\mu}\;. \label{tens}
\eeqa
The Hamiltonian operator follows from the ${\cal T}_{00}\equiv
{\cal H}$ component of the energy-momentum tensor in the form:
\beqa
{\rm H}&=&\int d^3 x\: {\cal T}_{00} \nonumber \\
&=&\int d^3 x \left\{ \psi^{\dag}
\left( -\vec{\alpha}\cdot \left( i\nabla + \vec{\Sigma}\den
\right)+ \beta (M-\Gamma_\phi\den \phi) +\Sigma_0 \right) \psi
-\bar\psi
\gamma^{\mu} \Sigma^{R}_{\mu}\psi
+\frac{1}{4}F_{\lambda\eta} F^{\lambda\eta}\right. \nonumber \\
&&\left.- \frac{1}{2} \left(\partial^\mu\phi\partial_\mu\phi -m_\phi^2
\phi^2\right)
+\frac{1}{4}\omega_{\lambda\eta}\omega^{\lambda\eta}-\frac{1}{2}
m_\omega^2\omega_{\lambda}\omega^{\lambda}
  +\frac{1}{4}\vec{\rho}_{\lambda\eta}\cdot
 \vec{\rho}^{\lambda\eta}-\frac{1}{2} m_\rho^2 \vec{\rho}_{\lambda}\cdot
\vec \rho^{\lambda} \right\}\;. \label{ham0}
\eeqa
Formally, the thermodynamics of the DDHFT follows from the
calculation of the grand partition function:
\beq
Z_G = Tr\;  \exp \left[{-\beta\left(H(\hat{\rho}) -
\sum_{i=p,n} \mu_i {\hat
B}_i \right)} \right] \equiv \exp\left[-\beta \Omega\right]\;,\label{GRAND}
\eeq
where the trace goes over a complete set of states in the
Fock space, $\beta$ is the inverse of the
temperature and  ${\hat B}_p$ and ${\hat B}_n$ correspond to the
proton and neutron number operator respectively, such that the
baryon number operator is $\hat B={\hat B}_p + {\hat B}_n$. Although
the equations of motion, eqs.(\ref{PHI}-\ref{DIRAC}), seem to be
similar to the usual QHD ones \cite{sw}, it should be noticed that
the source terms for the
meson fields contain in-medium correlations through the density
dependence of the vertices and  the rearrangement term contributes
explicitly for the baryon field equation. Of course the solution
of these field equations, as they stand, is a formidable task. For this
reason, we are forced to consider approximations in order to obtain a
tractable problem. Hence, we consider the DDHFT in the
mean field approximation for nuclear matter at finite temperature.
\section{DDHFT for nuclear matter in the mean field approximation}
We assume in this work, for the description of the nuclear matter,
static meson fields, rotational and translational symmetry, charge
conservation and no Coulomb field \cite{ringthimet}. Therefore, all
time and spatial derivatives vanish and only time-like components
of the $\phi$, $\omega$ and $\rho$ meson fields remain.  In the
mean field approximation, the meson fields are replaced by classic
condensed fields and the time-like components of the $\phi$,
$\omega$ and $\rho$ condensed meson fields are called $\phi_0$,
$\omega_0$ and $b_0$ respectively. In this approximation, only
$\bar \psi$ and
$\psi$ are taken to be quantized field operators. The meson fields
are obtained by thermal averages of the corresponding sources.
The involved approximations are discussed later. In the
nuclear rest frame, consistently with the MFA, we have
$\hat\rho=\bar\psi \gamma_0\psi$.  Under these approximations the
Hamiltonian of eq.(\ref{ham0}) reads:
\beqa
{\rm H}&=&\int d^3 x ~ {\cal H} \nonumber \\
&=&\int d^3 x \left\{ \psi^{\dag}
\left( -i\vec{\alpha}\cdot \nabla + \beta (M-\Gamma_\phi\den
\phi_0)
+\Sigma^{(0)}_0\den \right) \psi \right. \nonumber \\
&&\left. + \frac{1}{2} m_\phi^2 \phi_0^2 - \frac{1}{2}
m_\omega^2\omega_0^2 - \frac{1}{2} m_\rho^2 b_0^2
  \right\} \label{ham1}
\eeqa
where
\beqa
 \Sigma^{(0)}_0 = \Gamma_{\omega}\den \omega_0
+\frac{\Gamma_{\rho}\den}{2}\tau_3 b_0\;. \label{REARMFA}
\eeqa
It is important to emphasize that in the MFA the rearrangement
term cancels out in the energy density, as a consequence of
the cancellation of this term in the exact Hamiltonian of equation
(\ref{ham0}), since the
density dependent interaction terms in the Lagrangian density of equation
(\ref{lag1}) involves no derivatives with respect to $\hat \rho$.

Now we proceed to calculate the partition function. In order to perform
the trace calculation in (\ref{GRAND}), we follow the same procedure
used  in the calculation of quantum gases with density dependent
interactions \cite{GNS} and in the molecular field approximation
to the Heisenberg Hamiltonian \cite{Stanley}.
For the evaluation of the trace in eq.(\ref{GRAND}) it is necessary that,
in a convenient basis of states, the exponential function
decomposes itself in a sum of independent terms. As it is
seen next, this can be done only when we approximate the
Hamiltonian operator in eq.(\ref{ham1}) by a one-body operator,
obtaining a linear term in the baryon number operator. Analogously
to  ordinary statistical  mechanics problems \cite{GNS} we
take the Hamiltonian expanded around the equilibrium  mean density
$\rho_0=\langle\hat\rho\rangle$. We then keep only terms up to the
first order in $\hat{\rho}-\rho_0$. Higher order terms are going
to be neglected.

In the DDHFT it is assumed that the functional form for the density
dependent coupling operators is sufficiently well behaved as a function of
$\rho$. In
eqs.(\ref{ham1}-\ref{REARMFA}) we expand the vertices around $\rho_0$ in
the form,
\beqa
\Gamma_i\den = \Gamma_i(\rho_0)
+\left.\frac{\partial\Gamma_i}{\partial \hat\rho}\right|_{\hat\rho=\rho_0}
(\hat\rho -\rho_0)+\cdots \label{vertexp}
\eeqa
and substitute the normal ordered scalar and isovector density by
their thermal averages
\beqa
\bar \psi \psi &\rightarrow&  \langle \bar \psi \psi \rangle =
\rho_S ~, \nonumber \\
\bar \psi  \tau_3 \gamma_0 \psi &\rightarrow&  \langle \bar \psi
\tau_3 \gamma_0 \psi \rangle = \rho_3 \nonumber ~~.
\eeqa
In this way we obtain, for the Hamiltonian density up to first order in
($\hat\rho -\rho_0$):
\begin{eqnarray}
{\cal H} ={\cal H}(\rho_0) + \Sigma^R_0(\rho_0)
(\hat{\rho}-\rho_0) + \cdots \label{densexp}
\end{eqnarray}
where
\beqa
\Sigma^{R}_{0}(\rho_0)= \left.\frac{\partial
\Gamma_{\omega}\den}{\partial \hat{\rho}}\right|_{\hat\rho=\rho_0}
\omega_0 \rho_0 + \left.\frac{1}{2} \frac{\partial
\Gamma_\rho\den}{\partial \hat{\rho}}\right|_{\hat\rho=\rho_0}
b_0\rho_3
   - \left.\frac{\partial \Gamma_\phi\den}{\partial
\hat{\rho}}\right|_{\hat\rho=\rho_0} \phi_0\rho_S \;.\nonumber
\eeqa
To calculate the trace in eq.(\ref{GRAND}), we use the single
particle basis associated with the Dirac field equation,
eq.(\ref{DIRAC}), calculated at density $\rho_0$, where $\rho_0=
B/V$ is the equilibrium nuclear matter density. This equation can be
solved exactly if one seeks stationary
solutions of the plane wave form, $\psi=\psi ({\mathbf p}) \exp(i
{\mathbf p} \cdot {\mathbf x} -i \varepsilon({\mathbf p}) t)$,
where $\psi({\mathbf p})$ is a four component spinor. Such
procedure is well known in the literature and details can be found
in references \cite{sw,ringthimet}. So, in the MFA, the baryon
spinors are eigenvectors of the stationary Dirac equation:
\beqa
 \left( -i\vec{\alpha}\cdot
\nabla + \beta (M-\Gamma_\phi(\rho_0) \phi_0) +\Sigma_0(\rho_0)
\right) \psi = \varepsilon \psi\;,
\eeqa
and the baryon energy is given by:
\[
\varepsilon(p)\equiv \varepsilon^{(\pm)}=\Gamma_\omega
\omega_0+\frac{\Gamma_{\rho}}{2} \tau_3 b_0 + {\Sigma}^{R}_0 \pm
\left({\mathbf p}^2 + M^{\ast}\right)^{1/2}\;,
\]
where $M^*= M-\Gamma_\phi(\rho_0)\phi_0 $.  Going through the
usual steps to solve the Dirac equation, one finds that
$\varepsilon^{(+)}(\varepsilon^{(-)})$ corresponds to the baryon
$u(\mathbf p)$  (anti-baryon $v(\mathbf p)$) four component
spinor. Proceeding as in the ordinary Walecka model \cite{sw},
both positive and negative energy states $u$ and $v$ are found and
the field operator, in the Schr\"{o}dinger picture, can be
expanded as:
\beqa \psi = \sum_{\mathbf p \lambda} \left( u(\mathbf p ,\lambda)
~a_{\mathbf p \lambda}^{\dagger} + v(\mathbf p ,
\lambda)~b^{\dagger}_{\mathbf p \lambda} \right)~.\nonumber \eeqa
In the summation
$\lambda$ stands for the spin and isospin projections. The
operators $a^{\dagger}_{\mathbf p \lambda}$ and
$b^{\dagger}_{\mathbf p \lambda}$ are interpreted as creation
operators for baryons and anti-baryons and satisfy anticommutation
relations \cite{sw}. Thus, substituting the field operators into
eq.(\ref{ham1}) and taking into account the expansion given in
eq.(\ref{densexp})
 the normal ordered Hamiltonian takes the form:
\begin{eqnarray}
 {\rm H}&=&\sum_{\mathbf p \lambda}
 \left({\mathbf p}^2+M^{\ast 2}\right)^{1/2}
 \left(a^{\dagger}_{\mathbf p \lambda}
 a_{\mathbf p \lambda} +b^{\dagger}_{\mathbf p \lambda}
 b_{\mathbf p \lambda}\right)
 +\left(\Gamma_\omega \omega_0 + \Sigma^R_0(\rho_0)\right)
 \hat B +\frac{\Gamma_{\rho}}{2} b_0 {\hat B}_3  \nonumber \\
&&+V\left(\frac{1}{2} m_\phi^2 \phi_0^2 - \frac{1}{2} m_{\omega}^2
\omega_{0}^2 -\frac{1}{2} m_\rho^2 b_{0}^{2} - \Sigma^R_0(\rho_0)
\rho_0\right)\;,
\end{eqnarray}
where ${\hat B}_3 ={\hat B}_p - {\hat B}_n$, $V$ is the volume of
the system and the nucleon number operator is given by
\beqa
{\hat B}_i = \sum_{\mathbf p s}
\left(a^{\dagger}_{\mathbf p s\tau_3(i) } a_{\mathbf p s\tau_3(i)} -
 b^{\dagger}_{\mathbf p s\tau_3(i)} b_{\mathbf p
 s\tau_3(i)}\right)\;,\;\;\;\;\;\;i= p,n
 \nonumber\;.
 \eeqa
Since in the present approximation all operators in the
exponential function defining the partition function,
eq.(\ref{GRAND}), are diagonal in the basis of eigenstates of the
baryon and antibaryon number operators, the grand partition
function can be exactly calculated. The results are analogous to
those for the Walecka model \cite{sw} reading
\beqa
\Omega(\beta,\mu_p,\mu_n;\phi_0,\omega_0,b_0)&=&-\frac{1}{\beta} \ln Z_G
\nonumber \\
\Omega&=&V\left(\frac{1}{2} m_\phi^2 \phi_0^2 -
\frac{1}{2} m_{\omega}^2 \omega_{0}^2
-\frac{1}{2} m_\rho^2 b_{0}^{2} -\rho_0 \Sigma^{R}_0\right) \nonumber \\
&&-\frac{V}{\beta}\frac{2}{(2\pi)^3}\sum_{i=p,n}\sum_{\mathbf p}
\left[\ln\left(1+e^{-\beta(E^{\ast}({\mathbf
p})-\mu^{\ast}_i)}\right)\right.)\nonumber \\
&&\left.+\ln\left(1+e^{-\beta(E^{\ast}({\mathbf p})+\mu^{\ast}_i)}\right)
\right]\;, \label{omega}
\eeqa
where $E^{\ast}({\mathbf p})=\left({\mathbf p}^2+{M^*}^2\right)^{1/2}$ and
the effective chemical potentials, $\mu^{\ast}_p$ and
$\mu^{\ast}_n$, are defined as
\beqa
\mu^{\ast}_p&=&\mu_p-\Gamma_{\omega} \omega_0
-\frac{\Gamma_{\rho}}{2} b_0 -
\Sigma^{R}_0, \nonumber \\
\mu^{\ast}_n&=&\mu_n-\Gamma_{\omega} \omega_0
+\frac{\Gamma_{\rho}}{2} b_0 - \Sigma^{R}_0 \label{efchem}\;.
\eeqa
Notice the explicit contribution of the rearrangement term in the
above expressions. The pressure is obtained through the relation
$P=-\Omega/V$ and the energy density by:
\beqa
{\cal E}=\frac{E}{V}=\frac{1}{V} \frac{\partial (\beta
\Omega)}{\partial
 \beta}+ {\mu}_p {\rho}_p + {\mu}_n {\rho}_n\;.
\eeqa
The finite temperature meson field equations can be determined by
assuming that the thermodynamic potential $\Omega$ be stationary
for variations of these fields. This is to be expected for a
system in thermodynamical equilibrium. So,
\[
\frac{\partial \Omega}{\partial \xi}=0,\;\;\;\;\;\;\xi=\omega_0, b_0,\phi_0
\]
yield the coupled
equations
\beqa
m_\phi^2\phi_0
- \Gamma_\phi(\rho_0) \rho_s &=&0, \label{phi} \\
m_{\omega}^2 \omega_0
- \Gamma_{\omega}(\rho_0) \rho_0&=&0, \label{V0}\\
m_\rho^2 b_0 -\frac{\Gamma_\rho(\rho_0)}{2} \rho_3&=&0, \label{b0}
\eeqa
where the thermal scalar and baryonic densities are defined as
\beqa
\rho_s&=&\langle\bar \psi \psi\rangle = 2 \sum_{i=p,n} \int
\frac{\d^3p}{(2\pi)^3}
\frac{M^*}{E^{\ast}({\mathbf p})}\left(f_{i+}+f_{i-}\right), \\
\rho_0&=&\langle\bar \psi \gamma^0 \psi\rangle = \rho_p + \rho_n\;,
\label{rhoi} \\
\rho_3&=&\langle\bar \psi \gamma^0 \tau_3 \psi \rangle = \rho_p-\rho_n\;,
\label{rho3}
\eeqa
with
the distribution functions given by
\beqa
f_{i\pm}= \frac{1}{1+\exp[(E^{\ast}({\mathbf p})
\mp\mu^*_i)/T]}\;, \quad i=p,n \label{disfun}
\eeqa
and the proton and neutron densities $\rho_p$ and $\rho_n$ reading
\beqa
\rho_i =2\int\frac{\d^3p}{(2\pi)^3}(f_{i+}-f_{i-})\;,\quad i=p,n.
\label{denbary}
\eeqa
The equations of motion, eqs.(\ref{phi}-\ref{b0}), are completely
consistent with the MFA. They follow from
eqs.(\ref{PHI}-\ref{DIRAC}) when the sources are approximated by
their thermal averages as discussed above. For example, for the
scalar field source, $\Gamma_\phi(\hat\rho)\bar \psi \psi
\rightarrow \Gamma_\phi(\rho_0)\rho_S$, etc.

Next, we reobtain, within the Thomas-Fermi approximation, the
same expressions just obtained for the pressure, energy density
and meson fields, demonstrating the consistency of our
approximations.
\section{DDHFT in the Thomas-Fermi approximation}

We first define the thermodynamical potential, following the
notation in \cite{mp}, as
\beqa
\frac{\Omega}{V}= {\cal E} - T {\cal S} - \mu_p \rho_p - \mu_n \rho_n\;,
\label{Omega1}
\eeqa
where ${\cal E}$, ${\cal S}$ are the energy and entropy density
respectively, $T$ is the temperature, $\mu_p$ ($\mu_n$) is the
proton (neutron) chemical potential and $\rho_p$ and $\rho_n$ are
respectively the proton and neutron densities as given in
eq.(\ref{denbary}), calculated in such a way that
$\rho=\rho_p+\rho_n$. Here, the distribution functions $f_{i+}$
and $f_{i-}$ for particles and anti-particles have to be derived
in order to make the thermodynamic potential stationary for a
system in equili\-bri\-um. The entropy density $\cal S$ is obtained
from
\beqa
{\cal S}=-2\sum_{i=p,n}\int \frac{d^3p}{(2 \pi)^3} \left(
f_{i+} \ln \left(\frac{f_{i+}}{1-f_{i+}}\right) + \ln ({1-f_{i+}})
+ (f_{i+} \rightleftharpoons f_{i-})  \right)\;.
\eeqa
Next, we obtain the energy density in the TFA, where the conserved
energy-momentum tensor, eq.(\ref{tens}), becomes:
\beqa
{\cal T}^{\mu\nu}=\bar \psi i\gamma^\mu \partial^{\nu}\psi
-g^{\mu\nu}\left[- \frac{1}{2} m_\phi^2 \phi_0^2 + \frac{1}{2}
m_{\omega}^2 \omega_{0}^2 +\frac{1}{2} m_\rho^2 b_{0}^{2}+\bar\psi
\gamma_0 \Sigma^{R}_0 \psi \right]  \label{tem}\;.
\eeqa
>From the energy-momentum tensor one easily obtains the
Hamiltonian operator, which is the same as already displayed in equation
(\ref{ham1})
and from it, the energy density operator, which reads \cite{sw}:
\beqa
 {\cal E }&=&\frac{1}{V}\sum_{\mathbf p \lambda}
 \left({\mathbf p}^2+M^{\ast 2}\right)^{1/2}\left(
 a^{\dagger}_{\mathbf p \lambda}
 a_{\mathbf p \lambda} +b^{\dagger}_{\mathbf p \lambda}
 b_{\mathbf p \lambda}\right)
 +\Gamma_\omega \omega_0
 ~\frac{1}{V}\sum_{\mathbf p \lambda}\left(a^{\dagger}_{\mathbf p \lambda}
 a_{\mathbf p \lambda} - b^{\dagger}_{\mathbf p \lambda}
 b_{\mathbf p \lambda}\right)
 \nonumber \\
  &&+\frac{\Gamma_{\rho}}{2} b_0
\frac{1}{V}\sum_{\mathbf p s\tau_3} \tau_3\left(a^{\dagger}_{\mathbf
p s\tau_3}
 a_{\mathbf p s\tau_3} - b^{\dagger}_{\mathbf p s\tau_3} b_{\mathbf p
 s\tau_3}\right)
+\left( \frac{1}{2} m_\phi^2 \phi_0^2 - \frac{1}{2} m_{\omega}^2
\omega_{0}^2 -\frac{1}{2} m_\rho^2 b_{0}^{2}\right)\;, \label{enerdens}
\eeqa
where $\lambda =\{s, \tau_3\}$ are the spin and isospin indexes.
In obtaining the above equation we have assumed that the expectation
values of the meson fields are constant classical fields. Moreover the
Hamiltonian density is simply ${\cal H}(\rho_0)$, contrary to the
expression (\ref{densexp}) used in the calculation performed in the last
section.
The energy density in the TFA is obtained through the substitution of
the baryon and antibaryon number operators by their thermal
averages.
So, the energy density can be written in the semi-classical
TFA as
\beqa
{\cal E}&=& 2 \sum_{i=p,n} \int \frac{\d^3p}{(2\pi)^3}
\sqrt{{\mathbf p}^2+{M^*}^2} \left(f_{i+}+f_{i-}\right) +
\Gamma_{\omega} \omega_0 \rho_0 +\frac{\Gamma_{\rho}}{2} b_0
\rho_3
\nonumber \\
&&+\frac{m_\phi^2}{2} \phi_0^2-\frac{m_{\omega}^2}{2}
\omega_0^2 -\frac{m_{\rho}^2}{2} b_0^2\;.\label{enerd}
\eeqa
 After straightforward
substitutions, eq.(\ref{Omega1}) becomes
\beqa
\Omega&=& 2 \sum_{i=p,n} \int \frac{d^3p}{(2 \pi)^3}
\sqrt{{\mathbf p}^2+{M^*}^2} (f_{i+} + f_{i-}) + \Gamma_{\omega}
\omega_0 \rho +\frac{\Gamma_{\rho}}{2} b_0 \rho_3
+\frac{m_\phi^2}{2} \phi_0^2-\frac{m_{\omega}^2}{2} \omega_0^2
-\frac{m_{\rho}^2}{2} b_0^2\nonumber \\
&&+2 T \sum_{i=p,n} \int \frac{d^3p}{(2 \pi)^3} \left(
f_{i+} \ln \left(\frac{f_{i+}}{1-f_{i+}}\right) + \ln ({1-f_{i+}}) +
f_{i-} \ln \left(\frac{f_{i-}}{1-f_{i-}}\right) + \ln ({1-f_{i-}}) \right)
\nonumber \\
&&-2 \sum_{i=p,n} \int \frac{d^3p}{(2 \pi)^3}\:\mu_i (f_{i+}-f_{i-})\;.
\label{Omega2}
\eeqa
For a complete demonstration of the above shown expressions
obtained in a
Thomas-Fermi approximation for the non-linear Walecka model, please refer to
\cite{mp}.
At this point, eq.(\ref{Omega2}) is minimized in terms of the distribution
functions for fixed meson fields, i.e.,
\beqa
\left. \frac{\delta \Omega}{\delta f_{i+}} \right |_
{{f_{i-},f_{j\pm}, \phi_0,\omega_0,b_0}} =0 \quad i \ne j\;.
\eeqa
For the proton distribution function, the above calculation yields
\beqa
E^{\ast}({\mathbf p}) + \Sigma^R_0 - \mu_p + \Gamma_{\omega}
\omega_0 + \frac{\Gamma_{\rho}}{2} b_0 = -T\: \ln
\left(\frac{f_{p+}}{1-f_{p+}}\right).
\eeqa
Similar equations, with some sign differences are obtained for the
anti-proton, neutron and anti-neutron distribution functions. The
effective chemical potentials are the same as defined in
eq.(\ref{efchem}) and from the minimization we reobtain for the
distribution functions the expression given in eq.(\ref{disfun}).
In other words, with the distribution functions given in (\ref{disfun})
substituted into equation (\ref{Omega2}), equation(\ref{omega}) is exactly
reproduced. In the above calculation we have used $\rho_0$ and
 $\rho_3$ as defined in eqs.(\ref{rhoi}) and (\ref{rho3}) respectively.
 Within the Thomas-Fermi approach the
pressure becomes
\beqa
P&=&\frac{1}{3 \pi^2} \sum_{i=p,n}
\int \d p \frac{{\mathbf p}^4}{\sqrt{{\mathbf p}^2+{M^*}^2}} \left( f_{i+} + f_{i-}\right)
-\frac{m_\phi^2}{2} \phi_0^2 \left( 1 + 2 \frac{\rho}{\Gamma_\phi}
\frac{\partial
\Gamma_\phi}{\partial \rho} \right) \nonumber \\
&&+\frac{m_{\omega}^2}{2} \omega_0^2 \left( 1 + 2
\frac{\rho}{\Gamma_{\omega}} \frac{\partial
\Gamma_{\omega}}{\partial \rho} \right) +\frac{m_{\rho}^2}{2}
b_0^2 \left( 1 + 2 \frac{\rho}{\Gamma_{\rho}} \frac {\partial
\Gamma_{\rho}}{\partial \rho} \right)\;. \label{pressure}
\eeqa
Again we recover our previous expressions for the pressure and the
energy density. It is also important to stress that the
thermodynamical consistency, discussed in \cite{flw}, which requires
the equality of the
pressure calculated from the thermodynamical definition $\Omega=-PV$,
where
$\Omega$ is given in equation (\ref{omega}) or (\ref{Omega2}) and from
the energy-momentum tensor, is also
obeyed by the temperature dependent DDHFT. This
can be easily verified calculating the pressure from the thermal
average of the energy-momentum tensor (\ref{tem}) in the MFA as
\beqa
P=\frac{1}{3}\sum_{i=1}^3\langle {\cal T}_{ii} \rangle \;.
\eeqa

\section{Conclusions}

In this paper we have incorporated temperature effects in the DDHFT formalism
introduced in \cite{lf,tw}. Two different approximations were performed,
namely, a mean field approximation where the coupling operators are expanded in
powers of the density and the Thomas-Fermi approximation. In the first case,
the Hamiltonian is expanded in powers of $(\hat \rho - \rho_0)$ and the Dirac
equation is solved yielding the baryon energies. The grand partition function
is calculated and the thermodynamic quantities are derived. In the second case,
the thermodynamical potential is the starting point, where the energy density
is calculated in the semi-classical TFA. The thermodynamical potential is
minimized in terms of the unknown distribution functions, which are then
obtained. Thermodynamic consistency is proved.

As a final remark we discuss briefly another possible
application of the formalism we have derived above, i. e., the
study of effective chiral lagrangians with Brown-Rho (BR) scaling.
In 1991 Brown and Rho (BR) \cite{br} proposed an in-medium scaling law
for the masses and coupling constants for effective chiral
lagrangians. In ref.\cite{Song} the authors proposed an effective
Lagrangian whose parameters scale in nuclear medium according to
the BR scaling. In its simplest version the Lagrangian in the MFA
is \cite{Song,Song1}
\beqa
{\cal L} &=&\bar \psi\left[ \gamma_\mu\left(i\partial^{\mu}-
g^{\ast}_v(\rho) ~\omega^{\mu} \right)
-M^{\ast}(\rho) + h\phi \right]\psi \nonumber \\
&&+\frac{1}{2}\left(\partial_{\mu}\phi\partial^{\mu}\phi -m_\phi^{\ast
2}(\rho) \phi^2\right)
-\frac{1}{4}\omega_{\mu\nu}\omega^{\mu\nu}+\frac{1}{2}
m_\omega^{\ast 2}(\rho) \omega_{\mu}\omega^{\mu} \label{brlag} \;.
\eeqa
where in the notation of ref.\cite{Song1} $\psi$ is the
nucleon field, $\omega_\mu$ the isoscalar vector field, $\phi$ an
isoscalar scalar field and the masses with asterisk are BR-scaled
as introduced in ref.\cite{br}. The scaling of the vector coupling
constant is left arbitrary and $h$ is taken constant. So, the
Lagrangian in (\ref{brlag}) is of the form of a Walecka-like
Lagrangian and all the finite temperature formalism that we have
developed for the DDHFT can be immediately applied to these
lagrangians. The thermodynamics of effective lagrangians with BR
scaling has been studied in \cite{rho} for zero temperature. The
study of the validity of the BR scaling hypothesis for the
non-zero temperature case is currently under investigation.
\begin{center}
{\bf Acknowledgments}
\end{center}
This work was partially supported by CNPq - Brazil. We
would like to thank Dr. Marcelo Henrique Romano
Tragtenberg and Dr. Constan\c ca Provid\^encia
for very useful suggestions and productive discussions
related with this work.

\end{document}